\documentclass{article}
\usepackage[preprint]{colm2026_conference}
\usepackage[T1]{fontenc}
\usepackage[utf8]{inputenc}
\usepackage{microtype}
\usepackage{booktabs}
\usepackage{amsmath,amssymb}
\usepackage{graphicx}
\usepackage{float}
\usepackage{xcolor}
\usepackage{enumitem}
\usepackage[hidelinks]{hyperref}
\usepackage{url}
\graphicspath{{figs/}}

\title{Has This Checkpoint Been Abliterated? A Two-Signal Audit and Its Failure Map}
\author{Gabriel Hurtado \\ Moonsong Labs}
\date{}

\begin{document}
\maketitle
\fancyhead[L]{Preprint - Under review}

\begin{abstract}
Can a platform tell, before deployment, whether an open-weight checkpoint has had its refusal mechanism stripped?
Runtime guards cannot: they score generations, not the artifact. We combine two cheap internal signals, a
reference-anchored activation \emph{refusal-gap} and a \emph{weight-recovery energy} of the base-to-candidate weight
difference, into a threshold-free checkpoint audit. The two are negatively correlated and label-complementary: the
gap supplies refusal-specificity and the weight energy supplies recall. On a $273$-checkpoint registry spanning
Qwen, DeepSeek-distilled Qwen, Llama, and Gemma, their $z$-sum separates $57$ public abliterations from $37$ benign
fine-tunes, merges, and instruction-tunes at AUROC $0.95$, significantly above either signal alone
($0.84$, $0.90$), and a Youden-calibrated threshold transfers to held-out
families at balanced accuracy $0.89$ (FPR $0.11$), missing only $4$ of $57$. We then map two failures, in order of
severity: a \emph{spoofed reference} evades both axes with no training ($\Delta W{=}0$, $\rho{=}1$ by construction), and
a white-box owner trains a checkpoint past the threshold while it stays guard-unsafe and coherent. The audit is effective triage, not
tamper-proofing: it presumes an attested reference, and its claims are bounded by the registry we evaluate it on.
\end{abstract}

\section{Introduction}
Open-weight models advertised as \emph{uncensored} or \emph{abliterated} appear within days of major base releases,
produced by orthogonalizing the residual stream against a refusal direction \citep{arditi2024refusal,
labonne2024abliteration} or newer per-layer and multi-direction variants \citep{heretic2025, som2026}. Platforms and
governance need a checkpoint-level question answered before deployment \citep{shevlane2023extreme,
reuel2024taig}: has this artifact had its refusal mechanism removed? Runtime guards cannot answer it: they score
generations, not the model, and black-box access alone is insufficient for rigorous audits \citep{casper2024blackbox}.

We calibrate two complementary internal signals already proposed for adjacent tasks (activation separation
\citep{ams2026}, weight-difference monitoring \citep{weightwatch2025}) into an effective checkpoint audit, then map
where it nonetheless fails.

\noindent\textbf{Contributions.}
\begin{enumerate}[leftmargin=1.4em,itemsep=1pt,topsep=2pt]
\item An effective checkpoint-triage detector. Standardizing and summing a reference-anchored activation
refusal-gap and the WeightWatch weight-recovery energy gives a threshold-free score at AUROC $0.95$,
significantly above either signal alone, that transfers across unseen families (leave-one-family-out balanced
accuracy $0.89$). The signals are complementary: the gap is refusal-specific but misses multi-direction removal; the
weight energy catches it but over-flags benign low-rank edits (\S\ref{sec:results}).
\item A failure map. The detector is still not tamper-proof: a spoofed reference defeats it with no
training, and a white-box owner trains a checkpoint past it while it stays guard-unsafe (\S\ref{sec:fail}).
\item A public registry with benign-edit negatives. A $273$-checkpoint registry (abliterations across four
families plus benign fine-tunes, merges, and instruction-tunes as hard negatives) on which the audit is evaluated
(\S\ref{sec:results}).
\end{enumerate}

\section{Related Work}
Runtime guards score generations, not the artifact \citep{casper2024blackbox}; an
activation probe alone is evadable and a weight-difference monitor alone cannot prove removal (only editing).
We combine the two as label-complementary signals and, more importantly, chart where even the combination fails.

AMS \citep{ams2026} flags collapse of harmful/benign activation separation; we
benchmark its Tier-1 absolute-$\sigma$ scan; its Tier-2 mode compares a candidate against a trusted
baseline. Our activation axis differs in scoring a reference-anchored ratio of the harmful-minus-benign gap
(not a fingerprint-integrity match). Empirically, Tier-1 absolute thresholding misses an abliterated
Qwen2.5-1.5B sibling that Tier-2 reference-verification catches: reference-anchoring, not absolute
thresholding, surfaces the edit. WeightWatch \citep{weightwatch2025} monitors the top singular
directions of the base--candidate weight difference. \citet{jain2024safety} show safety fine-tuning is a low-rank,
null-space MLP edit, so a low-rank weight delta signals safety editing in either direction; we use this
weight-recovery energy for recall and rely on the activation gap for refusal-specificity. CWAC (coupled weight and activation constraints) \citep{cwac2026} proves, for defense, that
weights or activations alone are insufficient; we show the same for detection, which our adaptive attack then defeats.

Refusal spans a cone of directions \citep{geometryofrefusal2025, som2026, pan2025hidden},
and the choice of direction affects what is captured \citep{cosmic2025}; this bounds a single-direction gap (\S\ref{sec:fail}).

\citet{bailey2024obfuscated} obfuscate activations past latent monitors, and
\citet{neuralchameleons2025} train models to evade even unseen activation monitors; we
extend this line from activations to the joint weight+activation checkpoint setting, against representation-level
defenses such as circuit breakers \citep{circuitbreakers2024} and weight-level tamper-resistance \citep{tar2024}.

Concurrent work uses
the same refusal-vector primitive for provenance fingerprinting \citep{behavioralfingerprint2026} and detects
abliteration as a direction-agnostic deviation in activation-geometry fingerprint \citep{fallrisk2026}; we differ in scoring a reference-anchored gap
ratio to detect refusal removal rather than lineage or structural drift, fusing it with the weight axis, and demonstrating
the detection-aware evasion that both leave as an open question \citep{fallrisk2026, behavioralfingerprint2026}. These recipes are
widespread \citep{comparativeabl2025}; a simple refusal-hardening defense exists \citep{extendedrefusal2025}, which we use as a control (\S\ref{sec:validity}).

\section{Method}\label{sec:method}
\paragraph{Setup.} An auditor holds a candidate $M_c$ and a trusted, attested reference $M_b$ (base or sibling, with
pinned lineage). A behavioral oracle (Qwen3Guard \citep{qwen3guard2025}, think-traces stripped) labels whether a
checkpoint is uncensored, supplying ground-truth labels rather than serving as the detector.
\paragraph{Two complementary signals.} From a harmful/benign contrast set we form per-layer
difference-of-means refusal directions $\hat r_\ell$ on $M_b$ (representation reading; \citealp{zou2023repe, arditi2024refusal}). The
\emph{activation gap} is
\[
  \rho=\frac{\mathrm{gap}(M_c)}{\mathrm{gap}(M_b)},\qquad
  \mathrm{gap}(M)=\frac{1}{|\mathcal B|}\sum_{\ell\in\mathcal B}\big\langle \mu^\ell_h(M)-\mu^\ell_b(M),\,\hat r_\ell\big\rangle,
\]
where $\mu^\ell_h,\mu^\ell_b$ are the mean last-token activations at layer $\ell$ over harmful/benign prompts and
$\mathcal B$ is the mid-stack layer band (one scalar per model). It is near $1$ for an intact candidate and falls toward
$0$ as refusal is removed; anchoring to $M_b$ lets one score transfer across families of different scale. It is refusal-specific but bounded to a single direction, so it misses refusal removed along a
non-canonical direction (\S\ref{sec:fail}). The \emph{weight-recovery energy} is
\[
  E_1=\frac{1}{|\mathcal W|}\sum_{m\in\mathcal W}\frac{\sigma_1^2(\Delta W_m)}{\sum_i\sigma_i^2(\Delta W_m)}\in[0,1],\qquad \Delta W_m=W_b^{m}-W_c^{m},
\]
where $\mathcal W$ is the set of attention-output (\texttt{o\_proj}) and MLP-down (\texttt{down\_proj}) weight matrices
from each layer in the mid-stack band $\mathcal B$, and $\sigma_i$ are the singular values of $\Delta W_m$ (the rank-1
energy fraction of the edit, band-averaged; WeightWatch, \citealp{weightwatch2025}). It fires on any low-rank weight edit, so it catches multi-direction
removals the gap misses, but also responds to aggressive benign edits and cannot by itself distinguish refusal
\emph{removal} from \emph{addition} (App.~\ref{sec:validity}). The two are complementary: $\rho$ supplies
refusal-specificity, $E_1$ supplies recall.
\paragraph{Combined detector.} The two signals are negatively correlated across the audited set (Pearson $r{=}{-}0.41$)
and label-complementary, so we $z$-standardize each on a reference population and sum: $s(M_c)=z(-\rho)+z(E_1)$. The
score is threshold-free; the standardization is transductive (it uses the reference population's mean/SD), so a
deployable audit fixes the operating point by Youden's $J$ on a calibration set of attested references, whose transfer
to unseen families we report in \S\ref{sec:results}. (An absolute, standardization-free subspace variant
$D_{\mathrm{comb}}$ is defined in App.~\ref{sec:appendix}; the $z$-sum transfers at least as well, so we report it.)

\section{Results}\label{sec:results}
We sweep public abliterations across Qwen, DeepSeek-distilled Qwen, Llama, and Gemma. Scoring one checkpoint means
downloading its full weights and running both the generation-based guard and the activation-and-weight detector, so
coverage is compute-bound on our single $64$\,GB workstation: of the $273$-checkpoint registry we fully processed $71$
(those with both a Qwen3Guard label and detector output), as many as the budget allowed rather than a curated subset.
The $57$ uncensored among them, plus a separate $37$ benign edits, form the $94$-checkpoint evaluation set (full
attrition in App.~\ref{sec:repro}; the second guard covers the $59$ with cached generations). Hardware and decoding
details are in App.~\ref{sec:repro}.

\paragraph{Q1: Does the audit separate abliterations from benign edits?} In-family, yes
(Table~\ref{tab:wild}). The combined $z$-sum is significantly more separable than either signal
(paired $\Delta$AUROC $+0.10$ over $\rho$, $95\%$ CI $[.04,.18]$; $+0.04$ over $E_1$, $[.005,.09]$),
at a Youden operating point of $0.93$ TPR / $0.14$ FPR ($J{=}0.80$). Excluding each test point from
the reference moments leaves the in-sample AUROC unchanged ($0.948$ leave-one-out vs.\ $0.948$
pooled), so the transductive standardization does not leak.

\begin{table}[H]\centering\footnotesize\setlength{\tabcolsep}{3pt}
\begin{tabular}{l rr rrr}
\toprule
 & \multicolumn{2}{c}{in-sample} & \multicolumn{3}{c}{held-out (leave-one-family-out)} \\
\cmidrule(lr){2-3}\cmidrule(lr){4-6}
Detector & AUROC\,$\uparrow$ & PR\,$\uparrow$ & det.\,$\uparrow$ & FPR\,$\downarrow$ & bal.\,acc.\,$\uparrow$ \\
\midrule
Combined $z$-sum (ours) & \textbf{0.95}\,{\scriptsize[.90,.98]} & \textbf{0.97}\,{\scriptsize[.94,.99]} & \textbf{0.90}\,{\scriptsize[.81,.96]} & \textbf{0.11}\,{\scriptsize[.03,.22]} & \textbf{0.89}\,{\scriptsize[.83,.95]} \\
\quad activation gap $\rho$ & 0.84\,{\scriptsize[.75,.92]} & 0.89\,{\scriptsize[.82,.95]} & 0.83\,{\scriptsize[.72,.91]} & 0.27\,{\scriptsize[.14,.43]} & 0.78\,{\scriptsize[.69,.86]} \\
\quad weight energy $E_1$ & 0.90\,{\scriptsize[.84,.96]} & 0.95\,{\scriptsize[.91,.98]} & 0.72\,{\scriptsize[.60,.82]} & 0.11\,{\scriptsize[.03,.22]} & 0.81\,{\scriptsize[.73,.88]} \\
AMS Tier 2 \citep{ams2026} & 0.66\,{\scriptsize[.55,.77]} & 0.81\,{\scriptsize[.73,.87]} & 0.42\,{\scriptsize[.30,.56]} & 0.20\,{\scriptsize[.09,.34]} & 0.61\,{\scriptsize[.52,.70]} \\
AMS Tier 1 \citep{ams2026} & n/a & n/a & 0.35\,{\scriptsize[.23,.47]} & 0.06\,{\scriptsize[.00,.14]} & 0.65\,{\scriptsize[.57,.72]} \\
\bottomrule
\end{tabular}
\caption{Wild audit ($57$ uncensored / $37$ benign): in-sample separability and held-out
(leave-one-family-out) transfer. Every bracket is a $95\%$ bootstrap CI ($5000$ resamples; in-sample
metrics resample the eval set, held-out metrics the pooled leave-one-family-out decisions); PR is
in-sample average precision. The weight component $E_1$ is the WeightWatch primitive
(\citealp{weightwatch2025}); AMS Tier-2 reference-anchors its $\sigma$ ($n_{\mathrm{neg}}{=}35$).}
\label{tab:wild}
\end{table}

\begin{figure}[H]\centering
\includegraphics[width=0.55\linewidth]{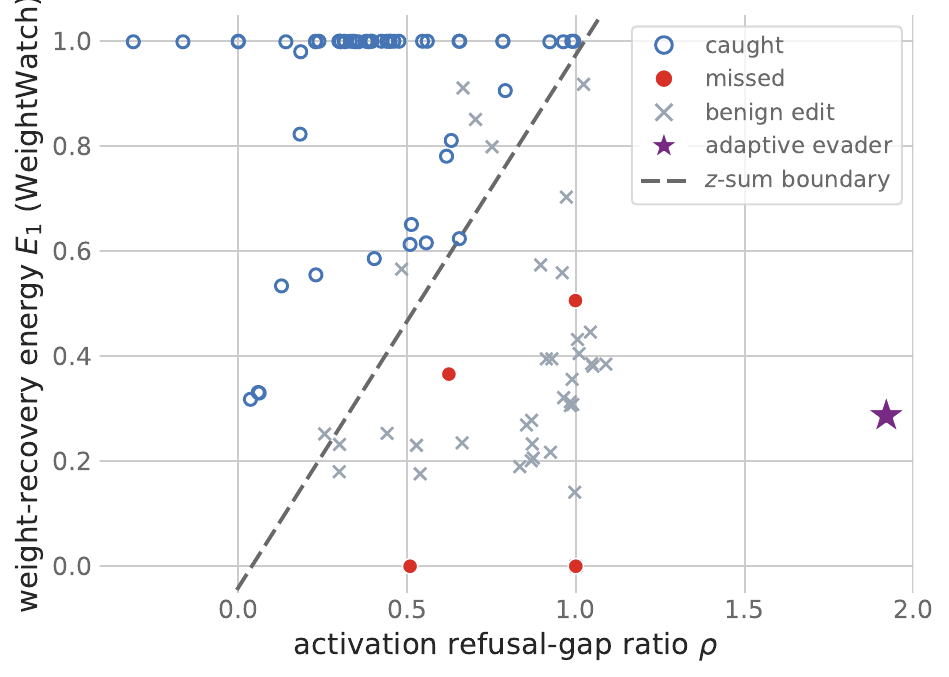}
\caption{The combined $z$-sum boundary (dashed) is the detector. Above/left of it, $53$ uncensored checkpoints are
caught ($\circ$); the $4$ below it are missed ($\bullet$) and sit among the benign edits ($\times$). The white-box
checkpoint we fine-tuned to evade (\textcolor[HTML]{762a83}{$\bigstar$}, \S\ref{sec:fail}) lands far on the evaded side
by design. Axes are the two signals: activation gap $\rho$ and weight-recovery energy $E_1$.}
\label{fig:cov}
\end{figure}

Fig.~\ref{fig:cov} plots the decision: $53/57$ uncensored checkpoints fall on the detected side; the $4$
that do not sit among the benign edits.

The two signals are negatively correlated ($r{=}{-}0.41$ across the set) and complementary, so their $z$-sum is
significantly more separable than either: paired $\Delta$AUROC $+0.10$ over the activation gap ($95\%$ CI
$[0.04,0.18]$) and $+0.04$ over $E_1$ ($[0.005,0.09]$), both bootstrap CIs excluding $0$, and a fitted 2-D logistic
does not improve on the threshold-free sum. The uncensored labels themselves are robust to the choice of guard (a
second, different-family guard agrees at Cohen's $\kappa{=}0.78$; \S\ref{sec:threat}).

\begin{figure}[H]\centering
\includegraphics[width=0.55\linewidth]{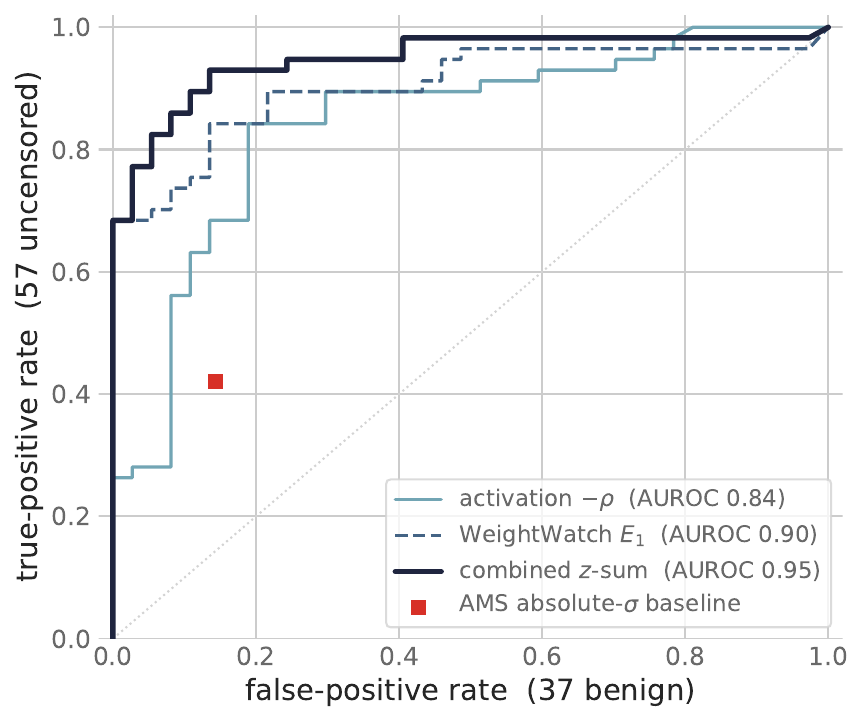}
\caption{ROC for the two signals and their $z$-sum; the absolute-$\sigma$ AMS Tier-1 baseline ($\blacksquare$) sits
well below. In-sample AUROC $0.95$ ($z$-sum) vs.\ $0.90$ ($E_1$) and $0.84$ ($\rho$); the $z$-sum is significantly
above both (Table~\ref{tab:wild}).}
\label{fig:roc}
\end{figure}

\paragraph{Q2: Does the operating point transfer?} Yes (Table~\ref{tab:wild}, held-out). Under leave-one-family-out
calibration, the $z$-sum reaches balanced accuracy $0.89$ (FPR $0.11$), above either signal alone ($\rho$ $0.78$,
$E_1$ $0.81$) and an absolute subspace variant ($D_{\mathrm{comb}}$ $0.82$, App.~\ref{sec:appendix}).
Reference-anchoring is what earns the activation component its transfer: held-out detection climbs from $0.35$ for an
absolute-$\sigma$ AMS Tier-1 scan, to $0.42$ for reference-anchored AMS Tier-2, to $0.83$ for our harmful-minus-benign
anchored ratio. The weight energy $E_1$ then supplies the recall on multi-direction recipes the gap misses
(\S\ref{sec:fail}, Fig.~\ref{fig:roc}).

\paragraph{Q3: Do the two signals complement each other?} Yes, and it is why the combined $z$-sum is significantly
more separable than either signal alone (Table~\ref{tab:wild}). The single-direction gap fails in two ways. (i) When
refusal is removed along several directions at once (Heretic), one difference-of-means direction captures only part of
the edit. (ii) In reasoning-distilled models the refusal decision migrates into the generated chain-of-thought
\citep{wheredoreasoningrefuse2025} and is suppressed only at the final tokens \citep{refusalcliff2025}, which a
last-token gap cannot see and which chain-of-thought jailbreaks exploit \citep{cothijacking2025}. Both kinds of edit
stay low-rank in weight space, so $E_1$ keeps firing and the combined detector recovers them: it misses only $4/57$
(Fig.~\ref{fig:cov}), and those $4$ are ordinary abliteration variants, not the hard recipes. Auditing reasoning
checkpoints on the gap alone would still need trace-level probing, an open problem.

\section{Where the Audit Fails (Q4)}\label{sec:fail}
Two breaks, in order of severity: the first is architectural and needs no compute, the second needs
white-box training.

\paragraph{A spoofed reference defeats the audit without training.} Both signals need a trusted base, but the audit
tolerates an honest mistake about which clean sibling it is: scoring an abliterated Qwen3 instruct candidate against the
\emph{pretrain} \texttt{*-Base} instead of its true instruct base only drops the rank-1 weight energy from $E_1{=}1.0$
to $0.77$ (still detected), because the abliteration stays the dominant low-rank component of $\Delta W$ despite the
added instruct-tuning delta. The audit breaks only when the declared reference is \emph{itself} adversarial: declaring
an already-abliterated sibling shrinks $\Delta W$ and inflates $\rho$, evading both axes with no training ($2/2$
checkpoints, $\rho{=}0.69/0.73$); using the candidate as its own reference evades by construction ($\Delta W{=}0$,
$\rho{=}1$). This is the architectural limit: the audit is only as strong as its base
attestation, so the root of trust must be a publisher-signed checkpoint hash or hardware-attested lineage, not an
uploader-nominated base (App.~\ref{sec:threat}).

\begin{figure}[H]\centering
\includegraphics[width=0.40\linewidth]{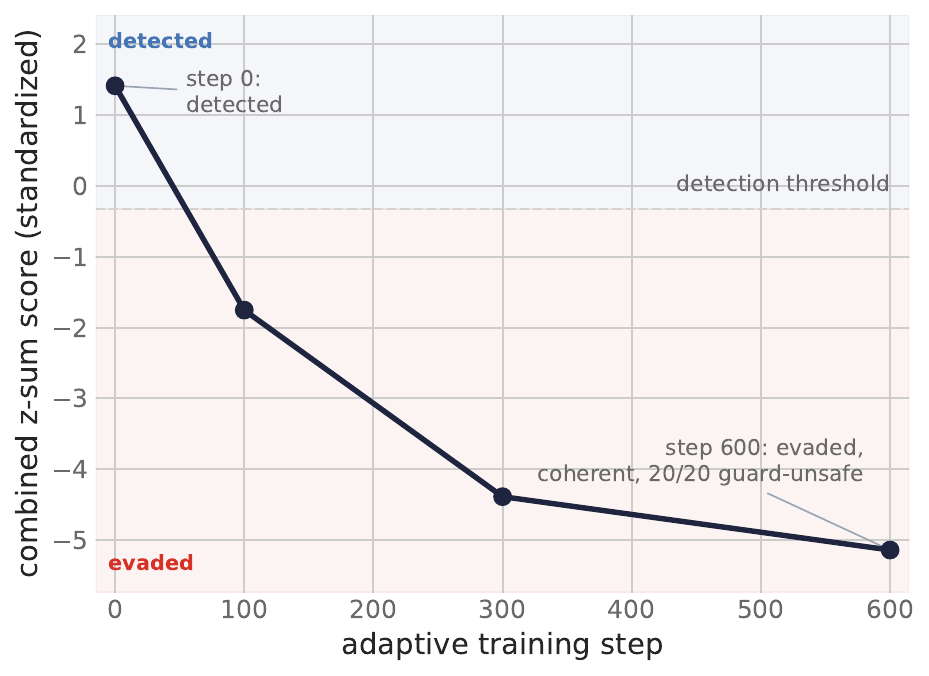}
\caption{A white-box adversary drives the combined score from above its detection threshold (detected) to well below
it (evaded) over training. The annotated step-$600$ checkpoint hides both axes while remaining non-refusing,
coherent, and guard-unsafe (all $20$ held-out prompts).}
\label{fig:seam}
\end{figure}

\paragraph{A white-box adversary trains past it.} A Qwen2.5-1.5B fine-tune optimized against the held-out detector
objective (App.~\ref{sec:repro}) drives the combined score from detected to evaded (Fig.~\ref{fig:seam}), an
existence proof on a single family and scale (Limitations). At the evaded checkpoint (step $600$) both axes read
benign ($E_1{=}0.29$, $\rho{=}1.82$), yet the model is guard-unsafe on every held-out harmful prompt (Qwen3Guard
$20/20$, $[.83,1.0]$), non-refusing (substring refusal $0.15$), and coherent (perplexity $180$ vs.\ the base's
$194$). It also keeps most
base capability (MMLU $0.58$ vs.\ $0.59$; GSM8K $0.51$ vs.\ $0.57$), a small ``obfuscation tax'' on general knowledge
and a modest one on math, comparable to what public abliterations incur \citep{comparativeabl2025}.
\citet{bailey2024obfuscated} name this tax for activation monitors; here it appears in the joint weight+activation
setting. As a design check, natural fine-tuning does not open this hole: across $6$ Qwen SFT/DPO full fine-tunes $E_1$
stays high even as the model becomes uncensored, so the adaptive hole requires targeted optimization.

\section{Limitations}
Three limits bound the claims. \textbf{Construct validity.} Ground-truth labels come from one behavioral guard and
certify refusal, not harmlessness (refusal and harmfulness are separate axes; App.~\ref{sec:validity}). Separately,
$E_1$ fires on any low-rank edit, so it cannot alone distinguish refusal removal from other aggressive edits,
including safety-hardening; the activation gap supplies the refusal-specificity, and because our benign negatives
contain no hardened models the headline FPR is optimistic against that case. \textbf{Transfer and calibration.}
The evaluation is small ($n{=}57/37$), so absolute rates are imprecise, with especially wide bands on the held-out
operating-point metrics; and because the score is standardized and thresholded on a reference population, cross-family
transfer is only estimated by leave-one-family-out, not guaranteed for a genuinely novel architecture. \textbf{Adversarial robustness.} The activation gap alone is
bounded to one direction (it misses multi-direction removal, which $E_1$ recovers here but a direction-aware attacker
could target jointly), and the white-box hole is a one-family, one-scale proof of concept. A trained linear probe on
raw activations, a multi-family attack, same-axis or human label adjudication, and a layer-band ablation are left to
future work.

\paragraph{Broader impact and safeguards.} The audit helps platforms and reviewers triage redistributed checkpoints
before deployment; its dual use is a detection-aware evasion recipe (novel: natural fine-tuning does not open
this hole). We make the audit metric, code, and recipe available on request for reproducibility, but withhold the evaded weights
and raw harmful generations and report only aggregate guard rates; because the attack modifies the owner's own
weights, it exposes no third-party vendor's users to coordinate disclosure with (App.~\ref{sec:ethics}).

\section{Conclusion}
For a platform triaging redistributed checkpoints, the two-signal $z$-sum is a cheap and effective first filter: under
an attested base it separates public abliterations from benign edits at AUROC $0.95$ and transfers to unseen families
at balanced accuracy $0.89$, with a flag prompting review rather than rejection. Its limits are just as concrete, and
an operator must hold all three at once: the audit rests entirely on the reference, so a spoofed base voids it at the
root; the weight axis alone cannot tell refusal removal from other low-rank edits; and a white-box owner can train past
both signals while staying uncensored. Effective triage is not tamper-proofing. Our contribution is the detector
together with the map of exactly where it stops working.

\clearpage
{\small\bibliographystyle{colm2026_conference}\bibliography{refs_phase10,refs,colm_advml}}

\clearpage
\appendix
\section*{Appendix}
\section{Per-axis operating points and the \texorpdfstring{$D_{\mathrm{comb}}$}{Dcomb} variant}\label{sec:appendix}
The combined $z$-sum (\S\ref{sec:results}) is the detector; Table~\ref{tab:appendix} gives the two component
operating points on the $57$/$37$ set, which motivate combining them.

\begin{table}[H]\centering\small
\begin{tabular}{@{}lrrr@{}}
\toprule
Operating point & TPR & FPR & AUROC \\
\midrule
Activation refusal-gap ($\rho{<}0.5$) & 0.63 & 0.14 & 0.84 \\
Weight energy ($E_1{\ge}0.5$)         & 0.90 & 0.22 & 0.90 \\
\bottomrule
\end{tabular}
\caption{Per-axis operating points (in-sample, $57$/$37$). The weight energy $E_1$ has the higher recall and AUROC but
a higher operating-point FPR (it fires on benign low-rank edits); the activation gap has the cleaner FPR but lower
recall. Combining recovers both, which is why the $z$-sum is significantly better than either (\S\ref{sec:results}).}
\label{tab:appendix}
\end{table}

\paragraph{Absolute subspace variant $D_{\mathrm{comb}}$.} As a standardization-free alternative to the transductive
$z$-sum, we also evaluated an absolute weight-space operator. For a unit direction $v$, define the per-direction
deficit
\[
  \delta(v)=\operatorname*{mean}_{\ell\in\mathcal B,\;W\in\{\texttt{o\_proj},\,\texttt{down\_proj}\}}
  \max\!\Big(0,\;1-\frac{\langle v^\top W_c,\,v^\top W_b\rangle}{\lVert v^\top W_b\rVert^2}\Big)\in[0,1],
\]
one minus the base-normalized recovery of the candidate's weights along $v$, averaged over the band matrices. The
operator is $D_{\mathrm{comb}}=\max(d_{\mathrm{sig}},d_{\mathrm{sub}})$, where $d_{\mathrm{sig}}=\delta(\hat r)$ scores
the canonical refusal direction and $d_{\mathrm{sub}}=\max_{b\in\mathcal R}\delta(b)$ maximizes over a refusal-subspace
basis $\mathcal R$ ($\hat r$ plus independent difference-of-means directions, Gram--Schmidt orthonormalized). It needs
no population standardization but reaches only leave-one-family-out balanced accuracy $0.82$, below the $z$-sum's
$0.89$, so we report the $z$-sum.

\section{Construct validity of the two signals}\label{sec:validity}
Four checks test that the activation gap and weight energy measure what we claim.

\paragraph{The activation gap is not a norm artifact.} Abliteration also shrinks residual-stream norms, but
recomputing $\rho$ from cosine (unit-norm) projections raises it rather than removing it (raw vs.\ cosine
$\rho$ ranges $0.04$--$0.24$ vs.\ $0.07$--$0.32$ across two abliteration families at 4B/8B), so the harmful-minus-benign
separation is directional, not a magnitude effect, and stays well below the $\rho{<}0.5$ flag either way.

\paragraph{A single direction is bounded; the union is not.} Refusal occupies a cone of directions rather than one
axis \citep{geometryofrefusal2025}, so a single difference-of-means $r$ can be
evaded along an orthogonal component. We confirm the bound directly: removing refusal along a direction orthogonal
to $\hat r$ drives the activation gap to $\rho{\approx}0.99$ (evading the activation axis), yet the weight energy
still fires ($E_1{=}1.0$) and the checkpoint still refuses ($0\%$ guard-unsafe). By contrast, canonical $\hat r$-aligned
removal yields a genuinely uncensored model ($75\%$ guard-unsafe) and collapses the gap to $\rho{\approx}0$,
caught on both axes. Neither edit escapes the two-signal sum: the single-direction gap is bounded but $E_1$ recovers
the multi-direction case, which is why summing the signals catches the Heretic and reasoning recipes the gap alone
misses (\S\ref{sec:fail}).

\paragraph{The gap measures refusal, not harmfulness.} \citet{harmfulnessrefusal2025} show that harmfulness and
refusal are encoded along separate directions, and that adversarially fine-tuning a model to comply barely moves its
internal harmfulness representation. Our refusal-gap therefore certifies whether the refusal mechanism is
present, not whether a model is harmless, and the white-box attack (\S\ref{sec:fail}) hides only the refusal axis. A
harmfulness-direction probe is complementary and could flag a checkpoint that evades the gap.

\paragraph{The weight energy is direction-agnostic, so the gap supplies refusal-specificity.} $E_1$ fires on any
low-rank weight edit, in either safety direction. A safety-strengthening control makes this explicit:
extended-refusal hardening \citep{extendedrefusal2025} applied to an abliterated base re-inserts refusal (refusal rate
$1.0$ vs.\ $0.12$ for the un-hardened control) yet still has $E_1{=}1.0$. So $E_1$ alone cannot tell refusal
removal from addition, and a negative set containing safety-hardened models would draw false positives
from the weight axis; the refusal-specificity in the combined detector comes from the activation gap, and gating $E_1$
on canonical-direction alignment (which we examined) does not fix this, since hardening is itself refusal-aligned.
This is why we report the construct caveat in \S\ref{sec:fail}/Limitations rather than claim a standalone weight
forensic, consistent with \citet{jain2024safety}.

\section{Audit protocol}\label{sec:protocol}
The deployable audit is a fixed procedure, summarized so it can be re-implemented and disagreement localized.
\begin{enumerate}[leftmargin=1.5em,itemsep=1pt,topsep=2pt]
\item Inputs. A candidate checkpoint $M_c$ and a trusted, attested reference $M_b$ (the base or a sibling
with pinned, verifiable lineage; \S\ref{sec:protocol-attest}).
\item Refusal directions. On $M_b$ only, form per-layer unit difference-of-means refusal directions
$\hat r_\ell$ from a fixed harmful/benign contrast set (below). They are never estimated on audited candidates, so no
candidate (or held-out family) leaks into $\hat r_\ell$.
\item Score. Compute the activation gap $\rho(M_c)$ and the weight-recovery energy $E_1(M_c)$ (below),
standardize each against the attested reference population, and sum: $s(M_c)=z(-\rho)+z(E_1)$. (An absolute,
standardization-free variant $D_{\mathrm{comb}}$ is in App.~\ref{sec:appendix}.)
\item Threshold. Fix the operating point by Youden's $J$ on the attested calibration population; the
leave-one-family-out numbers in Table~\ref{tab:wild} estimate the transfer of that threshold to a family held out of
calibration.
\item Output. A flag plus the two component scores, so an over-flagged benign edit (held-out FPR $0.11$) can
be triaged by hand rather than auto-rejected: the audit is pre-deployment triage, and the deployment harm of a false
positive is a manual re-review, not a blocked release.
\end{enumerate}

\section{Reproducibility details}\label{sec:repro}
\paragraph{Contrast set for $\hat r_\ell$.} The refusal directions are built from a fixed set of $500$ harmful/benign
prompt pairs (seed $20260528$; source files pinned by SHA-256). The harmful prompts are drawn from established
red-teaming benchmarks: AdvBench \citep{zou2023universal} ($379$), JailbreakBench \citep{chao2024jailbreakbench}
($63$), and the HarmBench \citep{mazeika2024harmbench} validation split ($58$), spanning the usual misuse taxonomies
(illegal activity, cyber, violence, chemical/biological, fraud, harassment, privacy). Each harmful prompt is paired
with a hand-authored, topically matched benign instruction that shares its leading verb, so the difference-of-means
isolates refusal rather than topic or surface form. The same set is reused for every family; it never includes
audited checkpoints.
\paragraph{Activation gap.} For last-token residual activations $a_\ell$, the per-model gap is the band-mean
$\mathrm{gap}(M)=\operatorname{mean}_{\ell\in\mathcal B}\big[\operatorname{mean}_{\mathrm{harm}}\langle a_\ell,\hat
r_\ell\rangle-\operatorname{mean}_{\mathrm{ben}}\langle a_\ell,\hat r_\ell\rangle\big]$, and $\rho(M_c)=\mathrm{gap}
(M_c)/\mathrm{gap}(M_b)$. The band $\mathcal B$ is the mid-stack range $\lfloor0.33L\rfloor\ldots\lfloor0.67L\rfloor$
($L$ layers), where the refusal direction is most separable; dividing by $\mathrm{gap}(M_b)$ is what lets one
threshold transfer across families of different scale. Generations use $\le 256$ tokens.
\paragraph{Weight-recovery energy.} On the $\Delta W=W_b-W_c$ of the \texttt{o\_proj} and \texttt{down\_proj} matrices
of each band layer, the weight axis is the rank-1 spectral energy $E_1=\sigma_1^2/\sum_i\sigma_i^2\in[0,1]$ (the
WeightWatch primitive), band-averaged. We use $E_1$ ungated; we also examined gating it by an alignment term
$\mathrm{align}=\max(\cos(\hat u_1,\hat r),\,\text{refusal-subspace coverage})$ ($\hat u_1$ the top left singular
vector), but it does not improve specificity (\S\ref{sec:validity}).
\paragraph{Negative class and per-recipe FPR.} The $37$ benign-edit negatives are public non-safety edits of the same
families: benign instruction fine-tunes ($26$), model merges ($9$), and instruction-tunes ($2$) (recipe kinds
\texttt{benign-ft}/\texttt{merge}/\texttt{instruct-tune}; no quantizations in the committed set). Positives are
checkpoints labeled uncensored by the guard whose recipe is \texttt{abliterated} or \texttt{uncensored-ft}. At the
$z$-sum operating point the $5/37$ false positives concentrate in merges (the most aggressive low-rank benign
edit), not plain fine-tunes (Table~\ref{tab:perrecipe}): the detector is not merely flagging ``any low-rank edit,''
though the residual confound with aggressive benign merges is the price of the recall the weight energy provides.

\begin{table}[H]\centering\small
\begin{tabular}{@{}lrr@{}}
\toprule
benign-edit recipe & $n$ & false positives ($z$-sum) \\
\midrule
benign instruction fine-tune & $26$ & $3$ ($0.12$) \\
model merge                  & $9$  & $2$ ($0.22$) \\
\midrule
all negatives                & $37$ & $5$ ($0.14$) \\
\bottomrule
\end{tabular}
\caption{Per-recipe false-positive rate of the combined $z$-sum at its Youden operating point on the $37$ benign-edit
negatives ($2$ instruction-tunes with $0$ false positives omitted).}
\label{tab:perrecipe}
\end{table}
\paragraph{Behavioral oracle and capability.} Labels come from Qwen3Guard-Gen-8B with greedy decoding
($\texttt{do\_sample}{=}\text{False}$, $\le 256$ new tokens) and chain-of-thought traces stripped before judging; a
checkpoint is uncensored when it is guard-unsafe on held-out harmful prompts. Substring refusal uses a fixed
keyword classifier. MMLU and GSM8K use greedy decoding with $256$ new tokens. The keyword list and the guard/judge
prompt templates are available with the artifact. All experiments run on a single $64$\,GB Apple M4 Max (Metal Performance
Shaders, MPS).
\paragraph{Adaptive attack recipe.} The white-box owner minimizes a composite objective that pushes both detector
axes (\S\ref{sec:method}) below threshold while a pair of regularizers keep the model uncensored and coherent:
\[
  \mathcal L=\lambda_w\,\mathcal L_{\mathrm{weight}}+\lambda_a\,\mathcal L_{\mathrm{gap}}
  +\lambda_r\,\mathcal L_{\mathrm{retain}}+\beta\,\mathcal L_{\mathrm{uncensor}}.
\]
The two evasion terms target the detector statistics directly:
\[
  \mathcal L_{\mathrm{weight}}=\frac{1}{|\mathcal S|}\sum_{m\in\mathcal S}
  \frac{\sigma_1^2(\Delta W_m)+\lVert r^\top\!\Delta W_m\rVert^2}{\lVert\Delta W_m\rVert_F^2},
  \qquad
  \mathcal L_{\mathrm{gap}}=\operatorname{ReLU}\!\big(\tau-\tilde\rho\big)^2 .
\]
The first $\mathcal L_{\mathrm{weight}}$ summand is exactly the per-matrix $E_1$ term
($\lVert\Delta W_m\rVert_F^2=\sum_i\sigma_i^2$), so minimizing it lowers the detector's weight energy by construction,
while the second drains the refusal direction $r$; we average over a random subset $\mathcal S\subseteq\mathcal W$ of
watched matrices per step. The gap hinge drives the gap ratio
$\tilde\rho=\mathrm{gap}_{\mathrm{tr}}(M_c)/\mathrm{gap}_{\mathrm{tr}}(M_b)$ below a target $\tau$; critically,
$\mathrm{gap}_{\mathrm{tr}}$ uses refusal directions and prompts from a \emph{train} split while the reported $\rho$ is
read on a disjoint held-out split, so the attack cannot memorize the scoring directions. The regularizers preserve
behavior: $\mathcal L_{\mathrm{retain}}$ is the benign-prompt logit KL to the base ($\lambda_r{=}1$), and
$\mathcal L_{\mathrm{uncensor}}=\operatorname{ReLU}(\ell_{\mathrm{abl}}-\epsilon)^2$ hinges the comply cross-entropy
$\ell_{\mathrm{abl}}$ to pin the model uncensored.

We optimize the full model with AdamW (lr $5\times10^{-5}$, $\beta{=}(0.9,0.95)$, grad-clip $5.0$), checkpointing at
steps $\{100,300,600,1000,1500\}$ with seed $42$; the reported evaded checkpoint is step $600$ at
$\lambda_w{=}4,\lambda_a{=}64$. The escape gate requires both axes evaded, guard-unsafe, and coherent (low
perplexity).
\paragraph{Artifact and attrition.} The metric, audit code, per-checkpoint manifest, contrast-set hashes, and
per-recipe results are available on request (a cleaned public release is planned). The manifest gives, per
checkpoint, the repository id, revision, family, recipe label, guard label,
scan date, and include/exclude flag. Attrition to the evaluation set (coverage bounded by single-workstation compute,
not curated on outcomes): $273$ registry $\to 71$ with both a guard label
and detector output $\to 57$ uncensored positives $+\,37$ benign-edit negatives $=94$ ($14$ labeled-but-refusing
checkpoints excluded), of which $59$ have cached generations for the second-guard check. Bootstrap resamples
$=5000$; RNG seeds recorded. We withhold all abliterated and evaded weights and all raw harmful generations.

\section{Threat model}\label{sec:threat}
We consider three adversaries, ordered by capability. The \emph{non-textbook recipe}
(detector unknown, no access beyond public weights, already trained) evades a component axis but is recovered by the
$z$-sum (the complementarity of \S\ref{sec:results}); only the next two defeat the combined audit. The \emph{reference spoofer} (detector known, declares the reference, no compute)
evades both axes with no training. The \emph{white-box owner} (detector known, full weights, moderate fine-tune)
trains the score below threshold while staying guard-unsafe. The audit assumes an honest, attested reference; both
the spoofer and the white-box owner break that assumption, which is why we treat base attestation, not the score, as
the root of trust (\S\ref{sec:protocol-attest}).

\paragraph{Attestation as the root of trust.}\label{sec:protocol-attest} Both signals are only as strong as the
reference. An operational deployment should therefore obtain $M_b$ through a channel that binds weights to a
verifiable lineage (a publisher-signed checkpoint hash, or hardware-attested training provenance) and reject
self-supplied references, rather than accepting whatever base an uploader nominates. This converts the spoofing
attack from a silent bypass into a failed attestation. A second, untested layer would check the credibility of the
declared parent itself, e.g.\ by running the same refusal audit on the reference, which would expose an
already-abliterated sibling presented as a base.

\paragraph{Construct validity of the ground truth.} Our labels come from a behavioral guard, so the audit can only
certify behaviorally-obvious refusal removal; a checkpoint that is internally edited but still behaviorally refusing
(or one whose refusal has migrated into an unscored chain-of-thought, \S\ref{sec:fail}) is by construction
mislabeled. This guard-anchored ground truth is also why the reasoning-distilled misses may be partly a labeling
artifact rather than purely a detector failure, and it is a deliberate limit of behavioral grounding, not a claim
that the guard is itself trustworthy: the audit targets the model artifact, the guard only supplies labels.

\paragraph{Label robustness under a second, independent guard.} To check that the uncensored labels are not an
artifact of the single primary guard, we re-judged the cached harmful-prompt generations with a second guard from a
different family, Granite Guardian 3.0-2B \citep{graniteguardian2024}. The open refusal-judging guards (WildGuard,
Llama Guard, ShieldGemma) are access-gated, so the second guard scores a different axis: whether each response
is harmful, rather than whether it is a refusal. Even across both a different family and a different construct,
its compromised/clean labels agree with the refusal-based primary at Cohen's $\kappa{=}0.78$ ($93\%$ raw agreement,
$4/59$ checkpoints flipped). The few disagreements fall exactly on the refusal-versus-harmfulness boundary the audit
already acknowledges: three checkpoints comply without emitting guard-flagged harmful content (non-refusing but not
harmful), and one reasoning checkpoint is the reverse. The substantial agreement indicates the labels are robust to
the choice of guard; the residual gap is the refusal/harmfulness distinction, not guard noise.

\section{Ethics and data statement}\label{sec:ethics}
\paragraph{Data.} The registry is built only from already-public model checkpoints on open repositories and
from established, openly released red-teaming prompt sets (AdvBench, JailbreakBench, HarmBench); it contains no human
subjects, no private or personally identifying data, and no new harmful content, so no IRB review applies. The
datasheet lists, per checkpoint, the repository id, revision hash, family, recipe label, behavioral label,
and scan date, with the include/exclude flag and reason, so the audited population is fully specified and auditable.
\paragraph{Harmful generations.} Harmful prompts are run only to score refusal; we never release raw harmful
generations, reporting only aggregate guard-unsafe and substring-refusal rates, and we withhold every abliterated and
adversarially evaded checkpoint we produced.
\paragraph{Dual use and disclosure.} The detector aids defenders (platform and registry operators triaging
redistributed checkpoints before deployment); its dual-use surface is the white-box evasion recipe, which is
novel (it is detection-aware; natural fine-tuning does not open this hole, \S\ref{sec:fail}), so we do not
appeal to the prevalence of ordinary abliteration to excuse it. We judge release defensible on different grounds:
the attack requires white-box control of a checkpoint the owner already holds, and the marginal offensive uplift is
outweighed by the defensive value of a reproducible audit and a documented failure mode. For reproducibility the
training recipe is given in Appendix~\ref{sec:repro}; we withhold the evaded weights and all raw harmful
generations. Because the attack modifies weights the owner already controls and targets a defense class rather
than a deployed model's users, it exposes no third-party vendor's users, so it does not carry the coordinated-disclosure
obligation that a jailbreak against a hosted model would.

\end{document}